\documentclass[10pt,twocolumn,letterpaper]{article}

\usepackage{cvpr}              
\usepackage[accsupp]{axessibility}

\usepackage{multirow}

\definecolor{cvprblue}{rgb}{0.21,0.49,0.74}
\usepackage[pagebackref,breaklinks,colorlinks,allcolors=cvprblue]{hyperref}

\def\paperID{31483} 
\def\confName{CVPR}
\def\confYear{2026}

\title{TM-BSN: Triangular-Masked Blind-Spot Network for Real-World Self-Supervised Image Denoising}

\author{Junyoung Park$^{1}$ \quad Youngjin Oh$^{1}$ \quad Nam Ik Cho$^{1,2}$\\
$^{1}$Department of ECE, INMC, Seoul National University, South Korea\\
$^{2}$IPAI, Seoul National University, South Korea\\
{\tt\small \{parkjun210, yjymoh0211, nicho\}@snu.ac.kr}
}

\begin{document}
\maketitle
\begin{abstract}
Blind-spot networks (BSNs) enable self-supervised image denoising by preventing access to the target pixel, allowing clean signal estimation without ground-truth supervision.
However, this approach assumes pixel-wise noise independence, which is violated in real-world sRGB images due to spatially correlated noise from the camera's image signal processing (ISP) pipeline.
While several methods employ downsampling to decorrelate noise, they alter noise statistics and limit the network's ability to utilize full contextual information.
In this paper, we propose the Triangular-Masked Blind-Spot Network (TM-BSN), a novel blind-spot architecture that accurately models the spatial correlation of real sRGB noise.
This correlation originates from demosaicing, where each pixel is reconstructed from neighboring samples with spatially decaying weights, resulting in a diamond-shaped pattern.
To align the receptive field with this geometry, we introduce a triangular-masked convolution that restricts the kernel to its upper-triangular region, creating a diamond-shaped blind spot at the original resolution.
This design excludes correlated pixels while fully leveraging uncorrelated context, eliminating the need for downsampling or post-processing.
Furthermore, we use knowledge distillation to transfer complementary knowledge from multiple blind-spot predictions into a lightweight U-Net, improving both accuracy and efficiency.
Extensive experiments on real-world benchmarks demonstrate that our method achieves state-of-the-art performance, significantly outperforming existing self-supervised approaches.
Our code is available at \url{https://github.com/parkjun210/TM-BSN}.
\end{abstract}    
\section{Introduction}
\label{sec:intro}

\begin{figure}[t]
    \centering
    \begin{subfigure}[b]{0.49\linewidth}
        \centering
        \includegraphics[width=\linewidth]{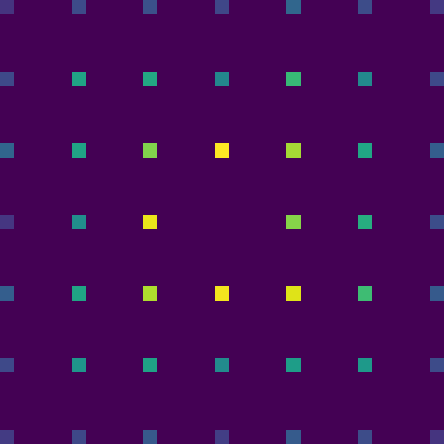}
        \caption{\centering\footnotesize AP-BSN~\cite{apbsn}}
    \end{subfigure}
    \hfill
    \begin{subfigure}[b]{0.49\linewidth}
        \centering
        \includegraphics[width=\linewidth]{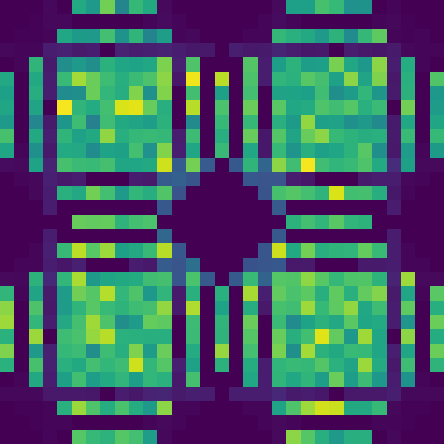}
        \caption{\centering\footnotesize LG-BPN~\cite{lgbpn}}
    \end{subfigure}
    \hfill
    \begin{subfigure}[b]{0.49\linewidth}
        \centering
        \includegraphics[width=\linewidth]{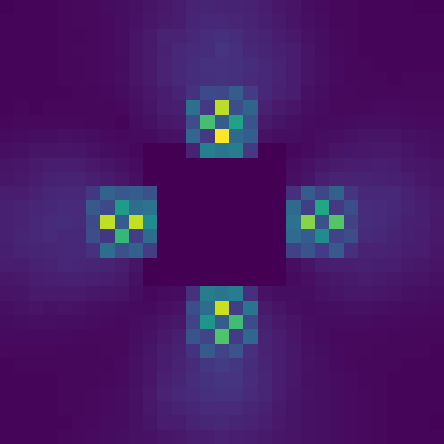}
        \caption{\centering\footnotesize AT-BSN~\cite{atbsn}}
    \end{subfigure}
    \hfill
    \begin{subfigure}[b]{0.49\linewidth}
        \centering
        \includegraphics[width=\linewidth]{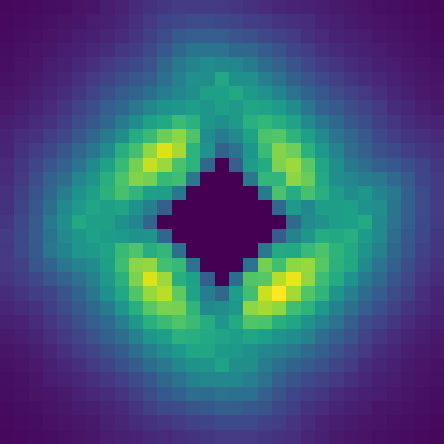}
        \caption{\centering\footnotesize Ours}
    \end{subfigure}
    \caption{Visualization of Effective Receptive Fields (ERFs) of different Blind-Spot Networks (BSNs): (a) AP-BSN, (b) LG-BPN, (c) AT-BSN, and (d) the proposed TM-BSN. The ERFs are computed by backpropagating the gradients from the central output pixel with respect to the input pixels~\cite{erf}.}
    \label{fig:fig1}
\end{figure}

Image denoising aims to restore a clean image from its noisy observation.
Traditional non-learning methods relied on explicit noise priors~\cite{tv} or nonlocal self-similarity~\cite{nlm, bm3d, wnnm}, but these methods often make strong assumptions about noise statistics and struggle to adapt to real-world settings.
With the advent of deep learning, image denoising research has moved toward data-driven approaches.
Early techniques used convolutional neural network (CNN) architectures trained on synthetic datasets, assuming simple noise models~\cite{dncnn, ffdnet}. They often achieved better results in removing synthetic noise and outperformed traditional algorithms.
However, they face challenges in generalizing to real-world camera images.
This struggle stems from the domain gap between synthetic and real noise, as real noise is usually more complex, signal-dependent, and spatially correlated.
Training on real noisy-clean image pairs~\cite{sidd, polyu} can help close this gap, but collecting such datasets is difficult, costly, and often impossible in dynamic or medical imaging scenarios.
These challenges have led to the development of self-supervised denoising methods that learn directly from noisy observations without needing clean images.

Among self-supervised frameworks, blind-spot networks (BSNs) are the leading method~\cite{hqssl, dbsn}.
By excluding the target pixel from the receptive field, BSNs prevent trivial identity mapping and enable the network to recover the clean signal from its surrounding context under the assumption of zero-mean, pixel-wise independent noise.
This design enables effective self-supervised learning using only noisy images.
However, this independence assumption does not hold for real-world sRGB images, where the demosaicing process in the camera’s image signal processing (ISP) pipeline creates strong spatial correlations among neighboring pixels.
These correlations violate the blind-spot learning principle, causing the network to reconstruct noise components instead of suppressing them.

To address this issue, several studies have used pixel-shuffle downsampling (PD) to decorrelate spatial correlations in noise~\cite{apbsn, puca, sdap}.
However, PD-based methods encounter a domain gap caused by differing sampling strides during training and inference.
Additionally, because inference is performed on downsampled images, they require extra post-processing to fix visual artifacts, increasing computational costs.
Recently, techniques operating at the original image resolution have been introduced to preserve textures and reduce the domain gap between training and inference~\cite{atbsn, aprrd}.
These methods typically employ tunable blind-spot strategies to expand the blind-spot region during training, thereby avoiding correlated noise.
While these methods result in better performance than PD-based approaches, they still cause unnecessary information loss by excluding potentially useful uncorrelated pixels.

In this paper, we introduce a new self-supervised denoising architecture based on the observation that the spatial correlation of sRGB noise exhibits a characteristic diamond-shaped pattern.
Such structured correlation arises from the demosaicing process, which reconstructs full-color images from the undersampled color channels of the camera’s color filter array (CFA).
To model this correlation geometry, we propose the Triangular-Masked Blind-Spot Network (TM-BSN), whose blind-spot is explicitly aligned with the spatial correlation structure of real-world sRGB noise.
TM-BSN uses a triangular-masked convolution that restricts the receptive field to its upper-triangular region, creating a diamond-shaped blind spot at the original image resolution, as shown in Fig.~\ref{fig:fig1}~(d).
This design effectively excludes pixels correlated with the target pixel while fully utilizing uncorrelated pixels, enabling accurate, detail-preserving denoising without downsampling or post-processing.
To further improve efficiency and harness knowledge across different blind-spot sizes, we adopt a knowledge distillation strategy that transfers information from multiple blind-spot predictions into a lightweight U-Net~\cite{unet}, enhancing both denoising performance and computational efficiency.

Our main contributions are summarized as follows:
\begin{itemize}
    \item We propose the Triangular-Masked Blind-Spot Network (TM-BSN), which employs a blind spot designed to match the spatial correlation geometry of real-world sRGB noise.
    \item We introduce a triangular-masked convolution that limits the receptive field to its upper-triangular region, allowing the formation of the diamond-shaped blind spot at the original image resolution without any downsampling.
    \item Extensive experiments on real-world benchmarks show that TM-BSN achieves state-of-the-art performance while maintaining fine texture details.
\end{itemize}

\section{Related Work}
\label{sec:Related_Work}

\subsection{Supervised Image Denoising}

CNN-based denoisers have greatly surpassed traditional optimization-based algorithms. Early supervised methods, such as DnCNN~\cite{dncnn}, FFDNet~\cite{ffdnet}, MemNet~\cite{memnet}, and RED30~\cite{red30}, were trained on paired noisy-clean datasets created with synthetic noise models (e.g., additive white Gaussian noise, AWGN).
While these models perform well on synthetic noisy images, they often struggle to generalize to real-world camera images due to a domain gap between synthetic and real-world noise.
To bridge this gap, CBDNet~\cite{cbdnet} explicitly modeled the image signal processing (ISP) pipeline to mimic the behavior of real camera noise.
Alternatively, several generative methods~\cite{gcbd,c2n,cycleisp,n2nf,neca} have been proposed to synthesize more realistic noise.

Despite these advances, supervised denoisers still struggle to handle the complex, signal-dependent, and spatially correlated noise generated by the ISP pipeline.
Although several large-scale real-world paired datasets, such as SIDD~\cite{sidd} and MIDD~\cite{midd}, have been introduced, their collection remains expensive, device-specific, and sometimes impossible, particularly in dynamic scenes or medical imaging.
These issues have driven the development of self-supervised denoising methods that do not require clean ground-truth data.

\subsection{Self-Supervised Image Denoising}

Instead of requiring paired noisy–clean data, self-supervised denoising methods are trained directly on noisy images.
Noise2Noise~\cite{n2n} first showed that a network can be trained using two independently corrupted images of the same scene.
Later, Noise2Void~\cite{n2v} and Noise2Self~\cite{n2s} introduced masking-based blind-spot learning, where the target pixel is masked during prediction to prevent trivial identity mapping.
These methods assume that noise is zero-mean and pixel-wise independent, allowing the network to estimate a pixel's clean signal from its surrounding context.
To improve masking-based training, HQ-SSL~\cite{hqssl} and DBSN~\cite{dbsn} proposed architecture-level blind-spot designs that enforce the blind spot structurally rather than through random masking.
This structural approach enhances stability and efficiency during training.
However, these methods are limited to synthetic, pixel-independent noise and perform less effectively on real-world camera images with spatially correlated noise.

To address this limitation, AP-BSN~\cite{apbsn} introduced an asymmetric pixel-shuffle downsampling (PD)~\cite{awgn} strategy that applies different PD stride factors during training and inference, effectively reducing spatial correlation while preserving structural fidelity.
This idea inspired several subsequent works, including PUCA~\cite{puca}, SDAP~\cite{sdap}, and TBSN~\cite{tbsn}, which further improved PD-based training by adding patch-unshuffle/shuffle, random subsampling, and transformer attention blocks.
Although these methods are effective at reducing noise correlation, they often sacrifice fine textures and require additional post-processing to eliminate checkerboard artifacts introduced by the PD operation.

Unlike approaches that rely on downsampling, AT-BSN~\cite{atbsn} works directly at the original image resolution by enforcing the blind-spot property through asymmetric operations, without using pixel-shuffle operations.
This design enables effective self-supervised learning while better preserving structural details.
Building on this foundation, APR-RD~\cite{aprrd} introduces the Adjacent Pixel Replacer (APR)—a sampling strategy inspired by Neighbor2Neighbor~\cite{nbr2nbr}—and the Recharged Distillation (RD) process, which improves the distillation stage.
However, these methods still rely on a rectangular blind-spot configuration, which does not align with the diamond-shaped correlation pattern inherent in the demosaicing of real sRGB images, thereby failing to fully utilize all uncorrelated pixels for effective denoising.

\section{Method}
\label{sec:method}

\subsection{Motivation}
\label{sec:motivation}

\begin{figure}[t]
    \centering
    \begin{subfigure}[t]{0.4\linewidth}
        \centering
        \includegraphics[width=\linewidth]{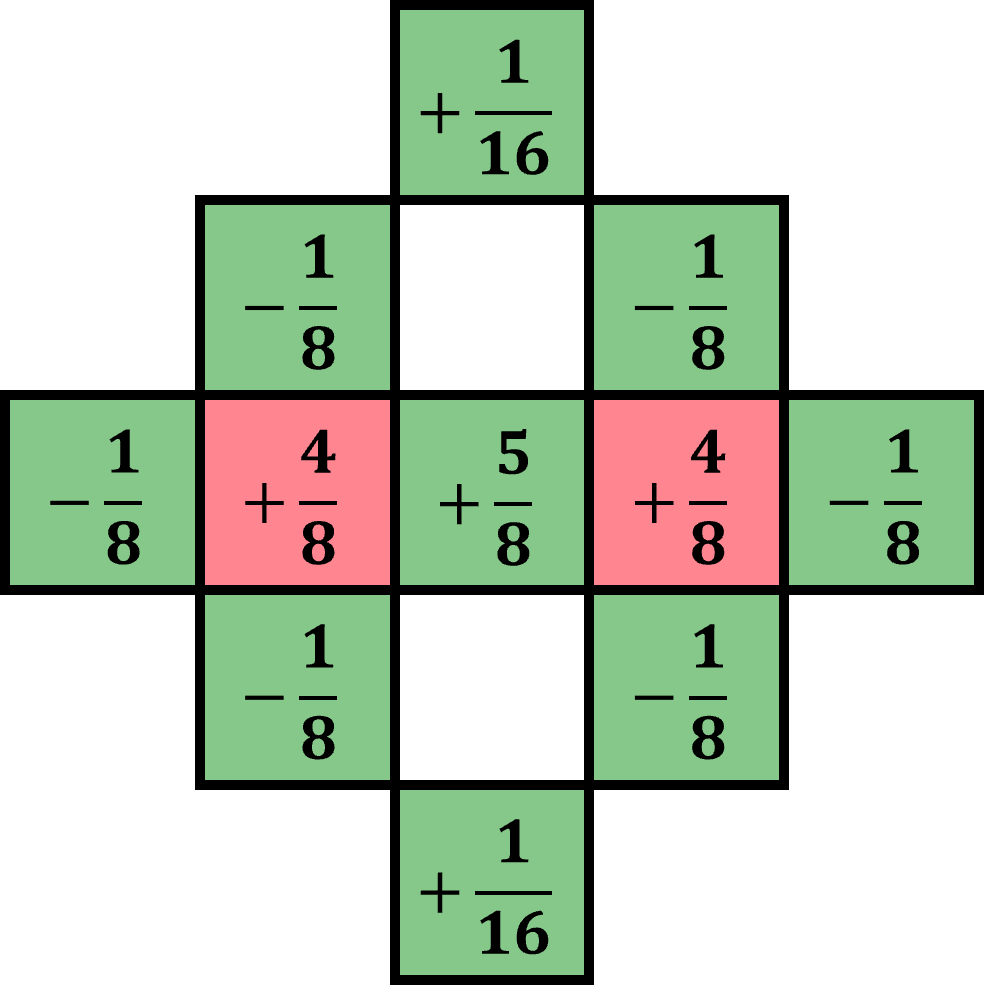}
        \caption{\centering\footnotesize Example of filter coefficients~\cite{malvar}.}
        \label{fig:fig2a}
    \end{subfigure}
    \hfill
    \begin{subfigure}[t]{0.55\linewidth}
        \centering
        \includegraphics[width=\linewidth]{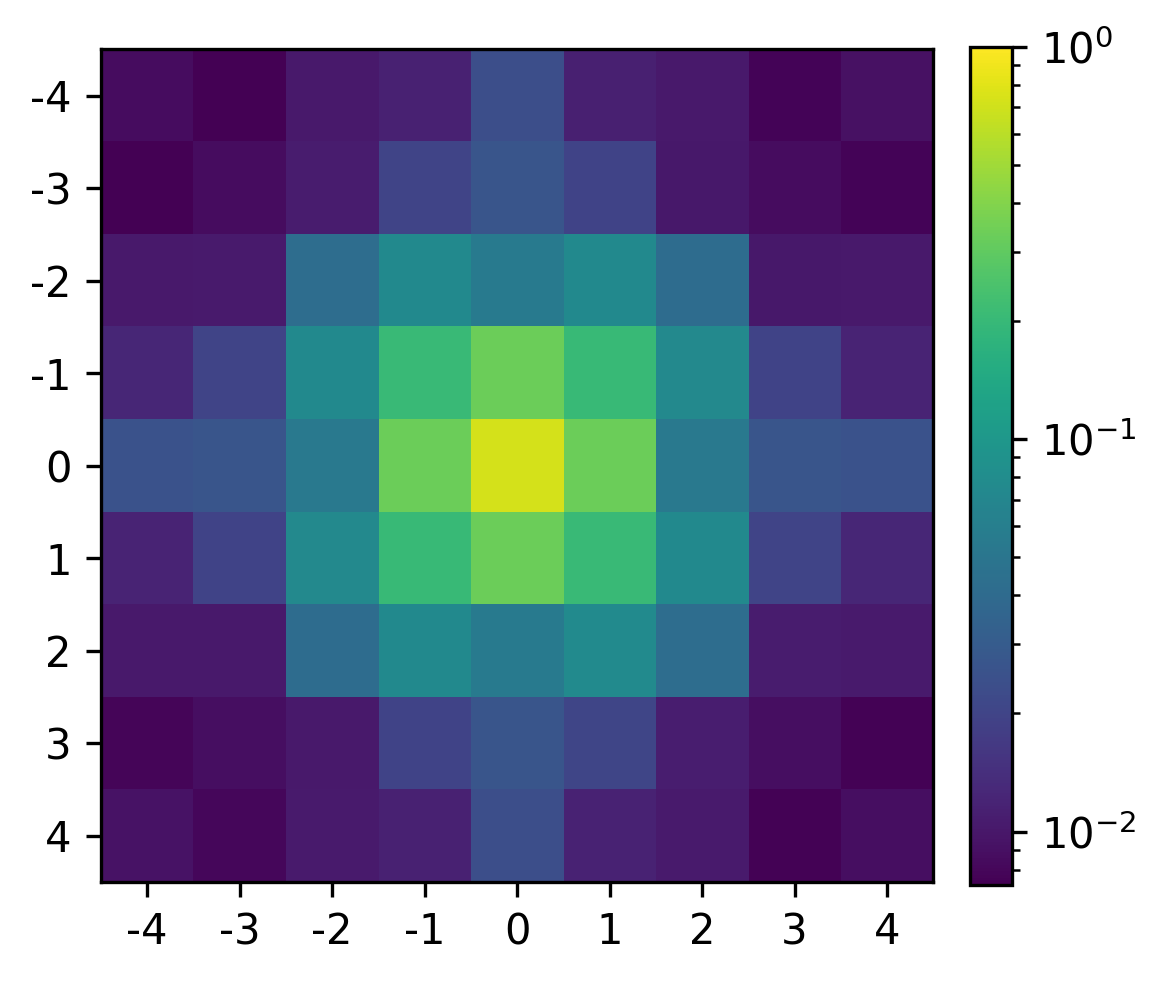}
        \caption{\centering\footnotesize Spatial correlation map of real noise measured from the SIDD dataset.}
        \label{fig:fig2b}
    \end{subfigure}

    \caption{(a) The demosaicing filter assigns higher weights to spatially closer samples during color reconstruction. 
    (b) Real noise exhibits a diamond-shaped correlation pattern with respect to relative distance.}
    \label{fig:fig2}
\end{figure}

During demosaicing, the missing color channels at each pixel are interpolated from neighboring samples captured by the camera’s color filter array (CFA).
Although numerous demosaicing algorithms have been proposed~\cite{malvar, pcsd, zhang, gunturk}, they generally share a common design principle: the interpolation relies on neighboring pixels, assigning higher weights to spatially closer samples while gradually reducing the influence of distant ones, as shown in Fig.~\ref{fig:fig2}~(a).
As a result, the reconstructed sRGB image exhibits a characteristic diamond-shaped spatial correlation pattern centered at each pixel, as depicted in Fig.~\ref{fig:fig2}~(b).
This spatially structured correlation violates the pixel-wise independence assumption, causing conventional BSNs to degenerate into near-identity mappings when trained on real noisy images.

\begin{figure*}[t]
    \centering
    \includegraphics[width=1\linewidth]{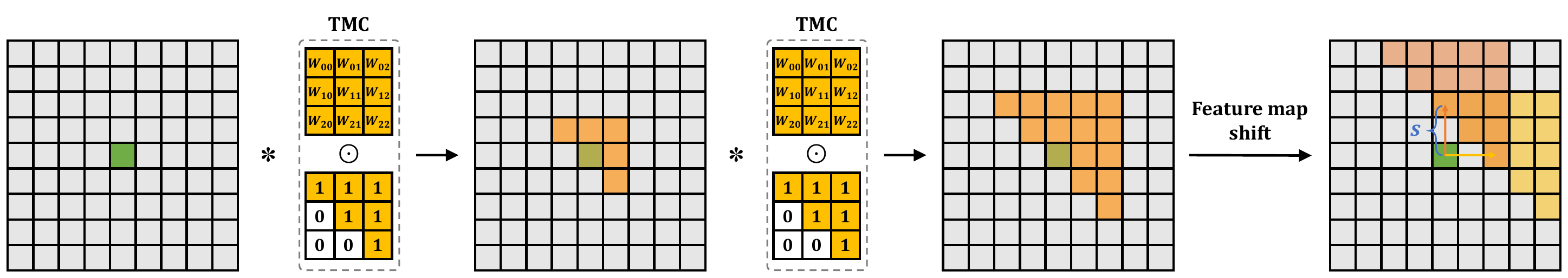}
    \caption{Illustration of receptive field expansion and blind-spot formation using the proposed triangular-masked convolution. The receptive field progressively expands in a triangular shape through stacked masked convolutions, and a feature-map shift is subsequently applied to introduce a central blind-spot.}
    \label{fig:fig3}
\end{figure*}
\begin{figure*}[t]
    \centering
    \includegraphics[width=1\linewidth]{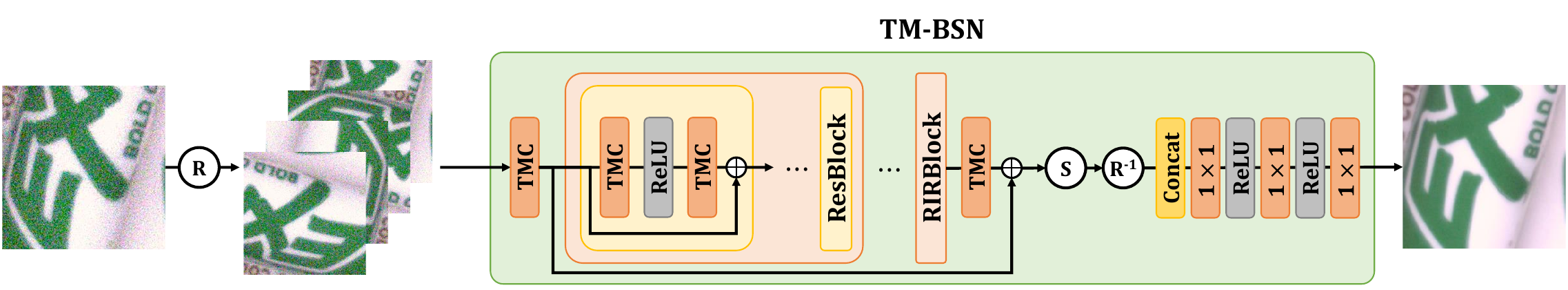}
    \caption{
    Overview of the proposed Triangular-Masked Blind-Spot Network (TM-BSN) architecture. 
    \textbf{R} rotates the input image by $0^\circ$, $90^\circ$, $180^\circ$, and $270^\circ$ and extracts features through TM-BSN; 
    \textbf{S} applies feature-map shifts by a given offset $s$ to generate blind spots; 
    and \textbf{R}$^{-1}$ unrotates the four branches, concatenates them along the channel dimension, and applies a $1{\times}1$ convolution to produce the final output.
    }
    \label{fig:fig4}
\end{figure*}

AP-BSN~\cite{apbsn} attempts to break this correlation by applying pixel-shuffle downsampling (PD)~\cite{awgn} with a large stride during training to decorrelate spatially correlated noise.
However, this strategy substantially limits the available receptive field, as shown in Fig.~\ref{fig:fig1}~(a).
LG-BPN~\cite{lgbpn} introduced a Densely-Sampled Patch-Masked Convolution (DSPMC) to mimic spatial correlation statistics, as illustrated in Fig.~\ref{fig:fig1}~(b).
While this method provides some benefit, the large kernel results in a high computational cost, and the blind region is not confined to the target pixel.
AT-BSN~\cite{atbsn} addresses these limitations by adopting feature-shift-based asymmetric operations, enabling self-supervised learning at the original image resolution.
This design creates a rectangular blind spot, as illustrated in Fig.~\ref{fig:fig1}~(c).
However, this rectangular configuration remains misaligned with the diamond-shaped spatial dependencies characteristic of demosaiced sRGB images, thereby excluding useful uncorrelated pixels near the corners.
To overcome these limitations, we propose TM-BSN (Triangular-Masked Blind-Spot Network), a novel blind-spot architecture that forms a diamond-shaped blind spot consistent with the spatial correlation geometry of real sRGB images, as shown in Fig.~\ref{fig:fig1}~(d).

\subsection{TM-BSN: Triangular-Masked Blind-Spot Network}
\label{sec:tmbsn}

To architecturally implement a blind spot that aligns with the diamond-shaped correlation pattern observed in real sRGB noise, we introduce the Triangular-Masked Convolution (TMC).
Unlike conventional convolutional operations, which use all kernel elements, TMC restricts its receptive field to the upper-triangular region of the kernel.
Formally, TMC is defined as:
\begin{equation}
F_{\text{out}} = (W \odot M) * F_{\text{in}} + b,
\quad
M_{ij} =
\begin{cases}
1, & i \le j, \\
0, & \text{otherwise},
\end{cases}
\end{equation}
where $F_{\text{in}}$ and $F_{\text{out}}$ denote the input and output feature maps, $W$ is a learnable $3\times3$ convolution kernel, $b$ is the bias term, and $M$ is a binary mask that zeros out the lower-triangular elements, with $i,j \in \{0, 1, 2\}$ being spatial indices within the kernel.
Here, $\odot$ denotes element-wise multiplication, and $*$ represents the standard convolution operation.
By stacking multiple TMC layers, we construct a network whose receptive field progressively expands across the upper-triangular area, as illustrated in Fig.~\ref{fig:fig3}.
This progressive expansion allows TM-BSN to capture contextual representations along the diagonal direction, which serves as the foundation for forming the diamond-shaped blind spot.

The overall architecture of TM-BSN is illustrated in Fig.~\ref{fig:fig4}. 
Building upon the AT-BSN architecture, TM-BSN adopts the restorer backbone from Variational Deep Image Restoration (VDIR)~\cite{vdir} to avoid pixel-level misalignment introduced by the pooling and un-pooling operations in U-Net~\cite{unet}. 
All \(3\times3\) convolution layers in the backbone are replaced with the proposed TMC layers, enabling the formation of the desired blind-spot structure at the original image resolution. 
The extracted feature maps are then adjusted using a feature-shifting operation, which enables a tunable blind-spot size controlled by the shift offset \(s\):
\begin{equation}
F_{\text{shift}}^{(s)} =
\text{concat}\big(
\text{Shift}_{\uparrow}(F, s),\,
\text{Shift}_{\rightarrow}(F, s)
\big).
\end{equation}
Here, \(\text{Shift}_{\uparrow}(F, s)\) and \(\text{Shift}_{\rightarrow}(F, s)\) denote vertical and horizontal shifting operations, respectively. 
Each shift operation applies zero-padding along the shifting direction and crops the opposite side to maintain spatial resolution, effectively displacing the receptive field of the target pixel by \(s\) pixels upward or rightward.
As illustrated in Fig.~\ref{fig:fig3}, this \(\text{Shift}(\cdot, s)\) operation excludes the target pixel from its own receptive field, thereby forming the blind spot.
We intentionally avoid diagonal shifts, as they double the stride, resulting in discontinuous spatial coverage.
Instead, concatenating the vertically and horizontally shifted features provides continuous blind-spot expansion along the diagonal axis.

Finally, TM-BSN aggregates feature maps from four rotated branches (0°, 90°, 180°, and 270°) to form a unified diamond-shaped blind spot.
This design effectively excludes spatially correlated regions introduced by the demosaicing process while fully leveraging uncorrelated contextual information.
As a result, TM-BSN preserves fine texture details and maintains signal and noise statistics of the original resolution.

\subsection{Knowledge Distillation}
\label{sec:distillation}

One of the key advantages of our feature-shifting architecture is its ability to efficiently generate multiple blind-spot predictions from shared feature representations. 
Since the feature extraction stage is shared across all shift offsets, TM-BSN can reuse the same intermediate features to produce several blind-spot outputs without redundant computation.
By varying the shift offset \(s\), TM-BSN generates multiple outputs corresponding to different blind-spot sizes with only about 15\% additional computational cost compared to a single forward pass.
These diverse blind-spot predictions provide complementary restoration cues that can be effectively distilled into a lightweight student network.

For knowledge transfer, we adopt the Recharged Distillation (RD) framework previously proposed in APR-RD~\cite{aprrd}. 
RD transfers complementary information from multiple blind-spot predictions by randomly re-injecting a subset of noisy pixels into each teacher output during training. 
The RD loss is defined as:
\begin{equation}
\mathcal{L}_{RD} = 
\sum_{s_i \in S}
\big\|
f_D(y) - \text{sg}\big[T_{s_i} \odot (1 - M_i) + y \odot M_i\big]
\big\|_1,
\label{eq:rd_loss}
\end{equation}
where \(y\) denotes the noisy input image, \(f_D(y)\) represents the student prediction, and \(T_{s_i} = \text{TM-BSN}(y; s_i)\) is the teacher output corresponding to the shift offset \(s_i\), while \(M_i\) is a random binary mask controlling the re-injected pixel subset. 
Here, \(S\) denotes the inference shift offset set, and \(\text{sg}(\cdot)\) indicates the stop-gradient operation, preventing gradient propagation to the pre-trained TM-BSN. 

The student network, implemented as a lightweight U-Net, is trained using the RD loss defined in Eq.~(\ref{eq:rd_loss}), and additional implementation details are provided in Sec.~\ref{sec:implementation}.
Through distillation, our method effectively enhances both denoising performance and computational efficiency. 
This improvement stems from transferring multiple complementary blind-spot predictions into a non-blind student network that can directly access target-pixel information.

\begin{table*}[t]
\centering
\caption{Quantitative comparison with state-of-the-art methods on the SIDD validation, SIDD benchmark, and DND benchmark.
We highlight the \textcolor{red}{best} and \textcolor{blue}{second-best} results among self-supervised methods.
$\dagger$ denotes results from a fully self-supervised setting on the DND benchmark (right side of the slash).
$\ddagger$ denotes results on the SIDD benchmark evaluated via the Kaggle competition using public checkpoints, as the official website is currently unavailable.}
\label{tab:table1}
\setlength{\tabcolsep}{5.2pt}
\renewcommand{\arraystretch}{1.12}
\begin{tabular}{lcccccc}
\toprule
\multirow{2}{*}{Methods} &
\multicolumn{2}{c}{SIDD Validation} &
\multicolumn{2}{c}{SIDD Benchmark} &
\multicolumn{2}{c}{DND Benchmark} \\
\cmidrule(lr){2-3}\cmidrule(lr){4-5}\cmidrule(lr){6-7}
& PSNR$^\uparrow$(dB) & SSIM$^\uparrow$ & PSNR$^\uparrow$(dB) & SSIM$^\uparrow$ & PSNR$^\uparrow$(dB) & SSIM$^\uparrow$ \\
\midrule
\multicolumn{7}{l}{\textbf{Non-Learning}}\\
BM3D~\cite{bm3d}     & 31.75 & 0.706 & 25.65 & 0.685 & 34.51 & 0.851 \\
WNNM~\cite{wnnm}     & 26.31 & 0.524 & 25.78 & 0.809 & 34.67 & 0.865 \\
\midrule
\multicolumn{7}{l}{\textbf{Supervised}}\\
DnCNN~\cite{dncnn}              & 37.73 & 0.943 & 37.61 & 0.941 & 37.90 & 0.943 \\
CBDNet~\cite{cbdnet}             & 33.07 & 0.863 & 33.28 & 0.868 & 38.05 & 0.942 \\
VDN~\cite{vdn}                 & 39.29 & 0.956 & 39.26 & 0.955 & 39.38 & 0.953 \\
Restormer~\cite{restormer}          & 39.93 & 0.960 & 40.02 & 0.960 & 40.03 & 0.956 \\
\midrule
\multicolumn{7}{l}{\textbf{Unpaired}}\\
GCBD~\cite{gcbd}                 &   \phantom{xx}-\phantom{xx}  &   \phantom{xx}-\phantom{xx}  &   \phantom{xx}-\phantom{xx}  &   \phantom{xx}-\phantom{xx}  & 35.58 & 0.922 \\
C2N~\cite{c2n}                 & 35.36 & 0.932 & 35.35 & 0.937 & 37.28 & 0.924 \\
\midrule
\multicolumn{7}{l}{\textbf{Self-Supervised}}\\
CVF-SID~\cite{cvfsid}                    & 34.51 & 0.941 & $35.03^{\ddagger}$ & $0.856^{\ddagger}$ &  $36.31$ / $36.50^{\dagger}$    &   $0.923$ / $0.924^{\dagger}$  \\
AP\mbox{-}BSN + R$^{3}$~\cite{apbsn}    & 35.76 &   \phantom{xx}-\phantom{xx}  & $36.51^{\ddagger}$ & $0.872^{\ddagger}$ &   \phantom{xx}-\phantom{xx} / $38.09^{\dagger}$      &   \phantom{xx}-\phantom{xx} / $0.937^{\dagger}$  \\
C\mbox{-}BSN~\cite{cbsn}                & 36.22 & 0.935 & $36.85^{\ddagger}$ & $0.882^{\ddagger}$     &   $38.45$ / $38.60^{\dagger}$   &   $0.939$ / $0.941^{\dagger}$   \\
LG\mbox{-}BPN~\cite{lgbpn}              & 37.32 & 0.886 & $37.67^{\ddagger}$ & $0.887^{\ddagger}$ & \phantom{xx}-\phantom{xx} / $38.43^{\dagger}$ & \phantom{xx}-\phantom{xx} / $0.942^{\dagger}$ \\
SDAP~\cite{sdap}                       & 37.30 & 0.894 & $37.64^{\ddagger}$ & $0.882^{\ddagger}$ & $37.86$ / $38.56^{\dagger}$ & $0.937$ / $0.940^{\dagger}$ \\
SASL~\cite{sasl}                  & 37.39 & 0.934 & $37.85^{\ddagger}$ & $0.879^{\ddagger}$ & $38.18$ / $38.58^{\dagger}$ & $0.938$ / $0.936^{\dagger}$ \\
PUCA~\cite{puca}                       & 37.49 & 0.880 & $37.93^{\ddagger}$ & $0.882^{\ddagger}$ & \phantom{xx}-\phantom{xx} / $38.83^{\dagger}$ & \phantom{xx}-\phantom{xx} / $0.942^{\dagger}$ \\
AT\mbox{-}BSN (D)~\cite{atbsn}          & 37.88 & 0.946 & $38.14^{\ddagger}$ & $0.891^{\ddagger}$ & $38.29$ / $38.68^{\dagger}$ & $0.939$ / $0.942^{\dagger}$ \\
APR (RD)~\cite{aprrd}                   & \textcolor{blue}{38.00} & \textcolor{blue}{0.947} & \textcolor{blue}{$38.26^{\ddagger}$} & \textcolor{blue}{$0.895^{\ddagger}$} & \textcolor{blue}{$38.57$} / $38.83^{\dagger}$ & \textcolor{blue}{$0.942$} / $0.944^{\dagger}$ \\
TBSN~\cite{tbsn}                       & 37.71 & \phantom{xx}-\phantom{xx} & $38.02^{\ddagger}$ & $0.886^{\ddagger}$ & \phantom{xx}-\phantom{xx} / \textcolor{blue}{$39.08^{\dagger}$} &  \phantom{xx}-\phantom{xx} / \textcolor{blue}{$0.945^{\dagger}$} \\
\textbf{\boldmath TM\mbox{-}BSN ($s{=}4$) (Ours)}   & 37.31 & 0.940 & $37.71^{\ddagger}$ & $0.881^{\ddagger}$ & 38.10 / $38.54^{\dagger}$ & 0.935 / $0.937^{\dagger}$ \\
\textbf{TM\mbox{-}BSN (D) (Ours)}     & \textcolor{red}{38.08} & \textcolor{red}{0.952} & \textcolor{red}{$38.31^{\ddagger}$} & \textcolor{red}{$0.900^{\ddagger}$} &  \textcolor{red}{$38.96$} / \textcolor{red}{$39.41^{\dagger}$} & \textcolor{red}{0.947} / \textcolor{red}{$0.949^{\dagger}$} \\
\bottomrule
\end{tabular}
\end{table*}

\section{Experiments}
\label{sec:experiments}

\subsection{Datasets}
We conduct experiments on two widely used real-world image denoising datasets: the Smartphone Image Denoising Dataset (SIDD)~\cite{sidd} and the Darmstadt Noise Dataset (DND)~\cite{dnd}.

\noindent\textbf{Smartphone Image Denoising Dataset (SIDD).}
SIDD is a large-scale dataset containing approximately 30,000 noisy images captured from ten scenes under various lighting conditions using five different smartphone cameras. 
For training, we use the SIDD Medium dataset, which consists of 320 noisy images. 
We evaluate our model on the SIDD validation and benchmark sets, each comprising 1,280 patches of size \(256\times256\).

\noindent\textbf{Darmstadt Noise Dataset (DND).}
The DND benchmark consists of 50 real-world noisy scenes captured by four consumer cameras. 
It provides 1,000 test patches in total, with 20 patches of size \(512\times512\) cropped from each scene. 
We evaluate our model on DND under two settings: (1) training on the SIDD Medium dataset and testing on DND, and (2) a fully self-supervised setting where the model is trained and evaluated directly on the DND dataset itself.

\subsection{Implementation Details}
\label{sec:implementation}

\noindent\textbf{TM-BSN training.} 
TM-BSN is trained using a self-supervised $L_1$ loss with a shift offset of \(s = 5\). 
The model is optimized using the Adam optimizer~\cite{adam} for 500k iterations and a batch size of 4. 
Training is conducted on randomly cropped patches of size \(128\times128\). 
The learning rate is initially set to \(1\times10^{-4}\) for the first 200k iterations and gradually decays to \(1\times10^{-6}\) over the remaining 300k iterations following a cosine annealing schedule.

\noindent\textbf{Knowledge distillation.} 
For knowledge distillation, we employ the U-Net~\cite{unet} (1.02M parameters) from AT-BSN~\cite{atbsn} to ensure a fair comparison with previous studies.
Distillation is performed using shift offsets \(S = \{2, 3, 4, 5, 6\}\), and the corresponding ablation analysis is provided in Sec.~\ref{sec:ablation}. 
The student network is trained for 200k iterations with a batch size of 8 on \(128\times128\) patches.
The learning rate is fixed at \(1\times10^{-4}\) for the first 100k iterations and then decays to \(1\times10^{-6}\) following a cosine schedule for the remaining 100k iterations.

\subsection{Comparison with State-of-the-Art Methods}

We compare the proposed TM-BSN with state-of-the-art self-supervised denoising methods, as well as representative non-learning, supervised, and unpaired approaches.
All self-supervised methods are trained on the SIDD Medium dataset for a fair comparison.
Quantitative evaluations are conducted using the standard PSNR and SSIM~\cite{ssim} metrics.

\noindent\textbf{Quantitative comparison.}
Since the official evaluation server for the SIDD benchmark is currently unavailable, all competing methods are evaluated using the publicly released checkpoints provided by their respective authors.
In this evaluation, SSIM is computed as the mean over the three RGB channels on the Kaggle competition, resulting in slightly lower absolute values compared to those reported by the official server.

Table~\ref{tab:table1} summarizes the quantitative results on the SIDD validation, SIDD benchmark, and DND benchmark datasets.
Our distilled TM-BSN (D) consistently achieves state-of-the-art performance across all three benchmarks.
Specifically, TM-BSN (D) improves on the previous best method, APR (RD)~\cite{aprrd}, by +0.05 dB on the SIDD benchmark, and surpasses TBSN~\cite{tbsn}, the prior best-performing approach on the DND benchmark, by +0.33 dB.

\begin{figure*}[h!]
	\newcommand{\siddsubfigurelen}{0.196}
	\centering
	\subfloat[\centering\scriptsize Noisy]
    {\includegraphics[width=\siddsubfigurelen\linewidth]{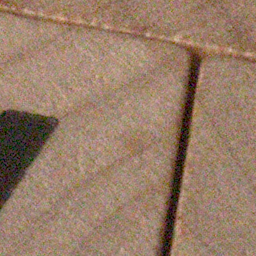}}
	\hfill
	\subfloat[\centering\scriptsize AP-BSN~\cite{apbsn}]{\includegraphics[width=\siddsubfigurelen\linewidth]{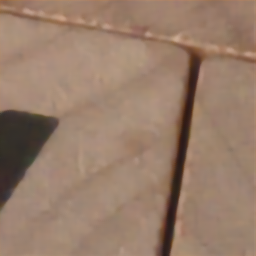}}
	\hfill
	\subfloat[\centering\scriptsize PUCA~\cite{puca}]{\includegraphics[width=\siddsubfigurelen\linewidth]{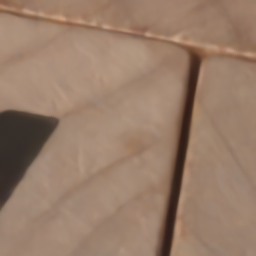}}
	\hfill
	\subfloat[\centering\scriptsize C-BSN~\cite{cbsn}]{\includegraphics[width=\siddsubfigurelen\linewidth]{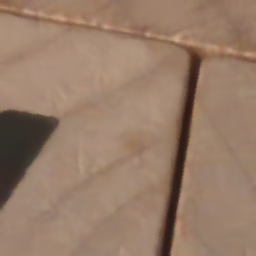}}
	\hfill
	\subfloat[\centering\scriptsize SASL~\cite{sasl}]{\includegraphics[width=\siddsubfigurelen\linewidth]{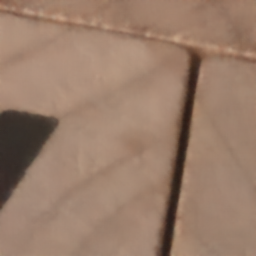}}\\
	\subfloat[\centering\scriptsize AT-BSN (D)~\cite{atbsn}]{\includegraphics[width=\siddsubfigurelen\linewidth]{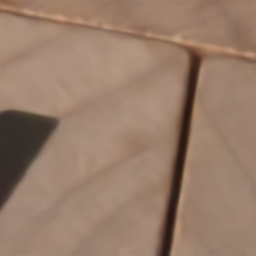}}
	\hfill
	\subfloat[\centering\scriptsize APR (RD)~\cite{aprrd}]{\includegraphics[width=\siddsubfigurelen\linewidth]{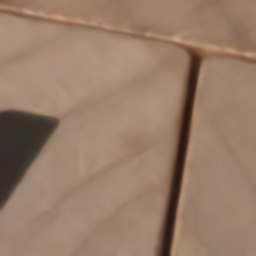}}
	\hfill
	\subfloat[\centering\scriptsize TBSN~\cite{tbsn}]{\includegraphics[width=\siddsubfigurelen\linewidth]{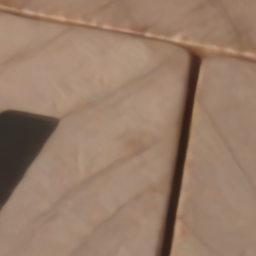}}
	\hfill
	\subfloat[\centering\scriptsize TM-BSN (D) (Ours)]
    {\includegraphics[width=\siddsubfigurelen\linewidth]{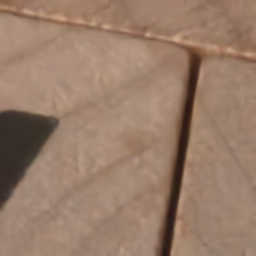}}
	\hfill
	\subfloat[\centering\scriptsize GT]
    {\includegraphics[width=\siddsubfigurelen\linewidth]{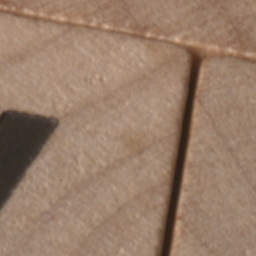}}
	\caption{Qualitative comparison on SIDD Validation dataset~\cite{sidd}.}
	\label{fig:sidd}
\end{figure*}

\begin{figure*}[h!]
	\newcommand{\siddsubfigurelen}{0.196}
	\centering
	\subfloat[\centering\scriptsize Noisy]
    {\includegraphics[width=\siddsubfigurelen\linewidth]{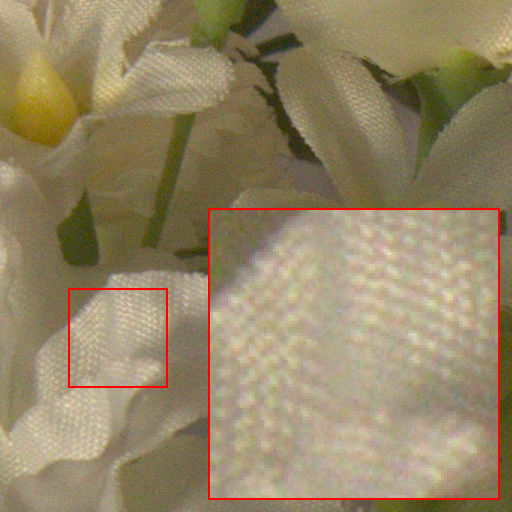}}
	\hfill
	\subfloat[\centering\scriptsize AP-BSN~\cite{apbsn}]{\includegraphics[width=\siddsubfigurelen\linewidth]{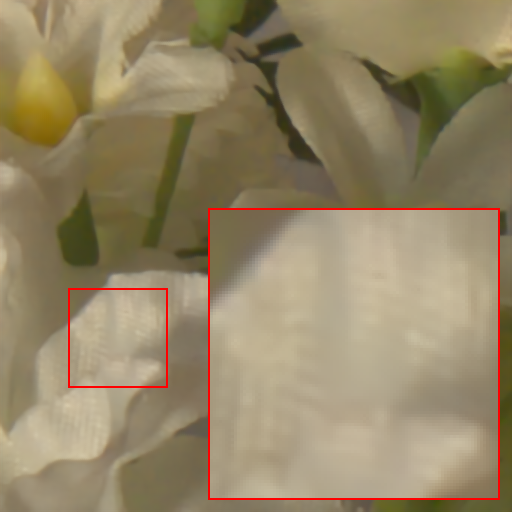}}
	\hfill
	\subfloat[\centering\scriptsize PUCA~\cite{puca}]{\includegraphics[width=\siddsubfigurelen\linewidth]{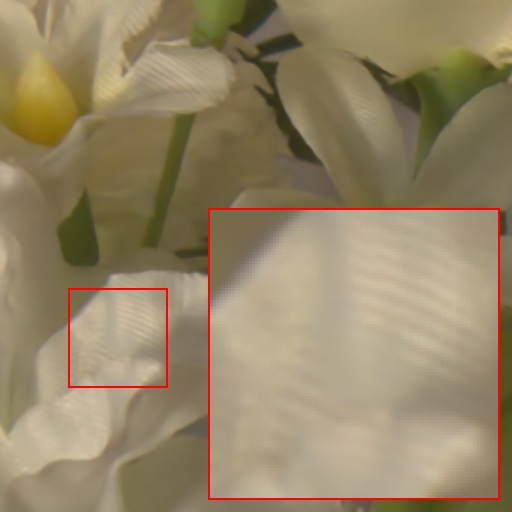}}
	\hfill
	\subfloat[\centering\scriptsize C-BSN~\cite{cbsn}]{\includegraphics[width=\siddsubfigurelen\linewidth]{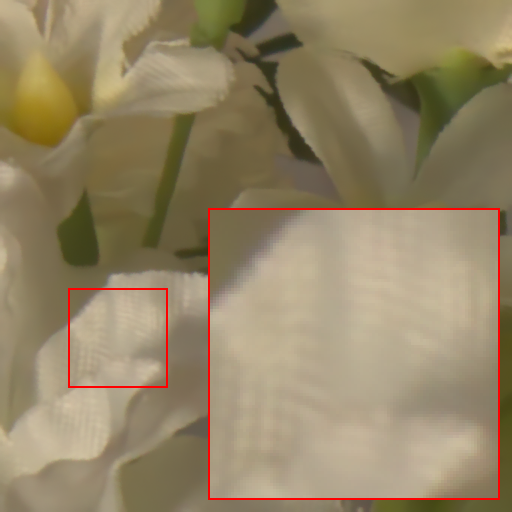}}
	\hfill
	\subfloat[\centering\scriptsize SASL~\cite{sasl}]{\includegraphics[width=\siddsubfigurelen\linewidth]{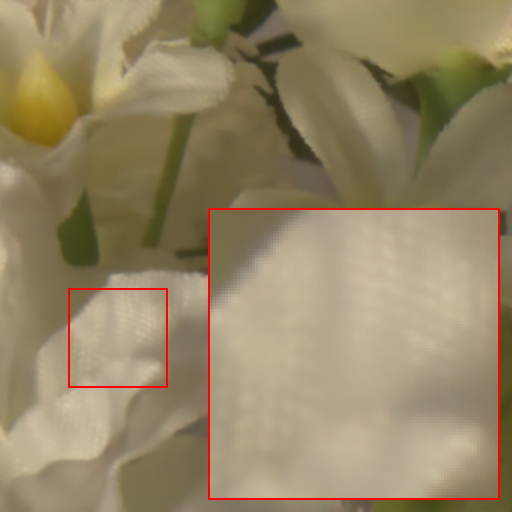}}\\
	\subfloat[\centering\scriptsize AT-BSN (D)~\cite{atbsn}]{\includegraphics[width=\siddsubfigurelen\linewidth]{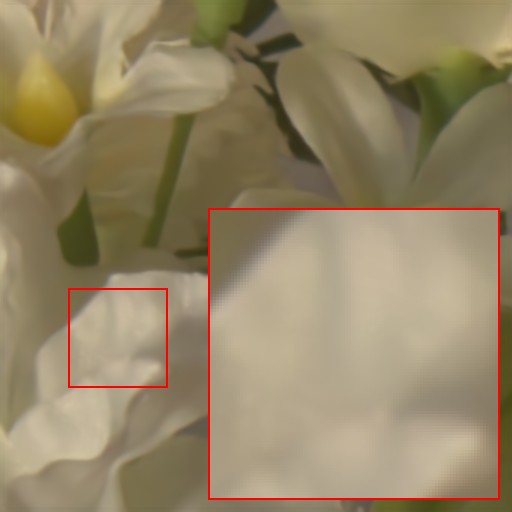}}
	\hfill
	\subfloat[\centering\scriptsize APR (RD)~\cite{aprrd}]{\includegraphics[width=\siddsubfigurelen\linewidth]{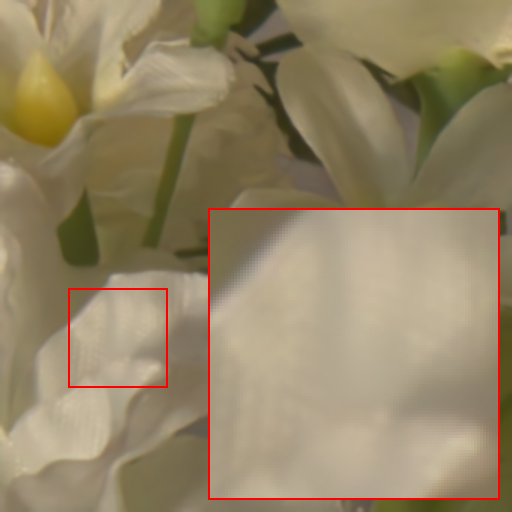}}
	\hfill
	\subfloat[\centering\scriptsize TBSN~\cite{tbsn}]{\includegraphics[width=\siddsubfigurelen\linewidth]{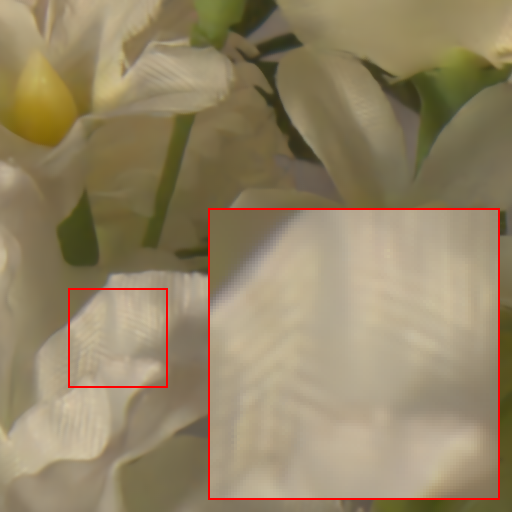}}
	\hfill
	\subfloat[\centering\scriptsize LG-BPN~\cite{lgbpn}]
    {\includegraphics[width=\siddsubfigurelen\linewidth]{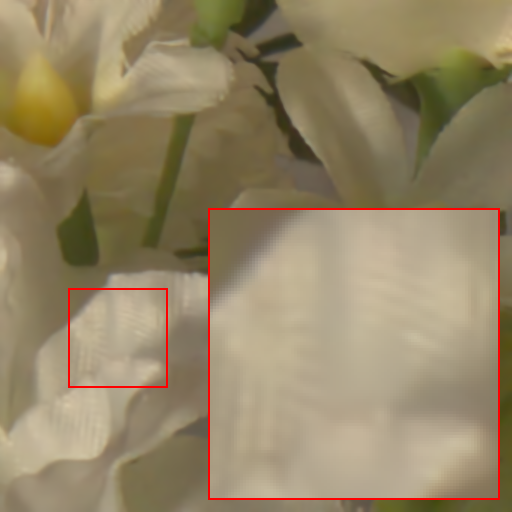}}
	\hfill
	\subfloat[\centering\scriptsize TM-BSN (D) (Ours)]
    {\includegraphics[width=\siddsubfigurelen\linewidth]{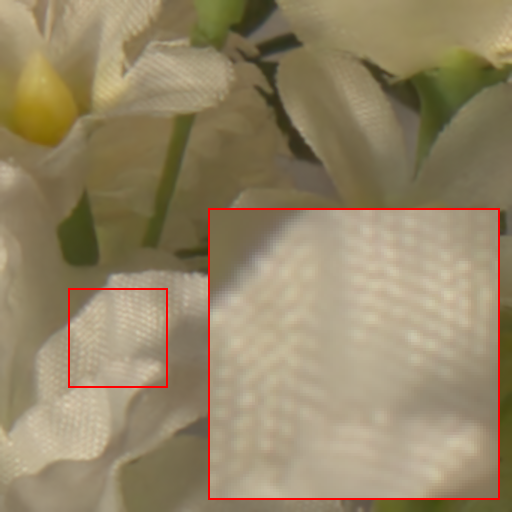}}
	\caption{Qualitative comparison on DND Benchmark dataset~\cite{dnd}.}
	\label{fig:dnd}
\end{figure*}

\noindent\textbf{Qualitative comparison.}
We provide qualitative comparisons with other self-supervised image denoising methods in Fig.~\ref{fig:sidd} and Fig.~\ref{fig:dnd}. 
As shown in Fig.~\ref{fig:sidd}, which depicts a wooden surface from the SIDD validation dataset, TM-BSN (D) effectively removes noise while faithfully preserving the fine wood grain and surface texture. 
In contrast, other methods tend to oversmooth the image, resulting in the loss of subtle structural details.
Furthermore, Fig.~\ref{fig:dnd} shows a challenging example of artificial flower petals from the DND benchmark. 
While competing methods suffer from excessive oversmoothing that removes delicate patterns, TM-BSN (D) robustly suppresses strong noise and preserves the grid-like fabric texture. 
These results highlight the superior ability of our method to distinguish fine textures from noise.

\begin{table}[h]
\centering
\caption{Comparison of model complexity and performance on the SIDD validation dataset. 
FLOPs and runtime are measured on \(256\times256\) input patches using an NVIDIA GeForce RTX~3090 GPU.}
\label{tab:table2}
\setlength{\tabcolsep}{3pt}
\renewcommand{\arraystretch}{1.15}
\begin{tabular}{lrrrr}
\toprule
\multirow{2}{*}{Method} &
\multicolumn{1}{c}{Params$^\downarrow$} &
\multicolumn{1}{c}{FLOPs$^\downarrow$} &
\multicolumn{1}{c}{Runtime$^\downarrow$} &
\multicolumn{1}{c}{PSNR$^\uparrow$} \\
& \multicolumn{1}{c}{(M)} &
  \multicolumn{1}{c}{(G)} &
  \multicolumn{1}{c}{(ms)} &
  \multicolumn{1}{c}{(dB)} \\
\midrule
AP-BSN~\cite{apbsn} & \hphantom{0}3.66 & 4456.76 & 287.32 & 35.76 \\
PUCA~\cite{puca}          & 12.78            & 2637.37 & 429.11 & 37.49 \\
C-BSN~\cite{cbsn}         & \hphantom{0}3.66 &  481.54 & 34.49 & 36.22 \\
SDAP (E)~\cite{sdap}      & \hphantom{0}3.66 &  956.85 & 51.51 & 37.30 \\
SASL~\cite{sasl}    & \hphantom{0}1.08 &   34.93 & 4.60 & 37.39 \\
AT-BSN~\cite{atbsn}        & \hphantom{0}1.27 &  164.60 & 14.84 & 36.80 \\
TBSN~\cite{tbsn}         & 12.97            & 5463.90 & 1004.60 & 37.71 \\
TM-BSN        & \hphantom{0}1.35 &  633.69 & 61.40 & 37.31 \\
\midrule
TM-BSN (D)    & \hphantom{0}1.02 &   26.74 & 3.21 & 38.08 \\
\bottomrule
\end{tabular}
\end{table}

\noindent\textbf{Complexity comparison.}
We present a comprehensive comparison of model complexity and inference speed on the SIDD validation set, as summarized in Table~\ref{tab:table2}. 
AP-BSN~\cite{apbsn}, PUCA~\cite{puca}, and TBSN~\cite{tbsn} require considerable computational overhead due to the random-replacing refinement process~\cite{apbsn}, which acts as a post-processing step for performance enhancement. 
In contrast, TM-BSN achieves competitive performance in a single forward pass, maintaining a favorable balance among parameters, FLOPs, and runtime efficiency. 
Furthermore, the distilled TM-BSN (D) improves both accuracy and efficiency, ultimately achieving state-of-the-art performance among self-supervised denoising methods.

\begin{figure}[h]
    \centering
    \includegraphics[width=1\linewidth]{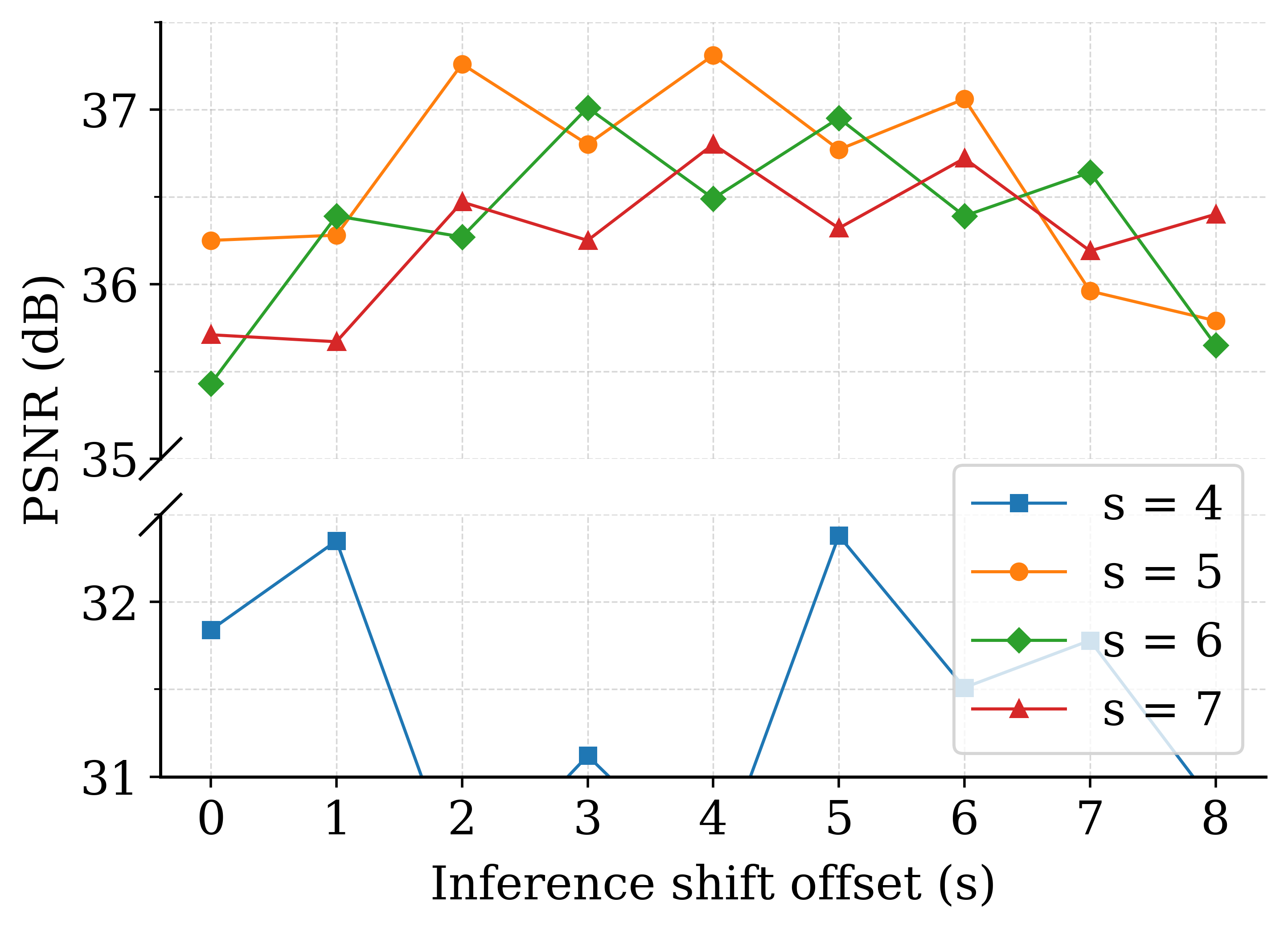}
    \caption{Ablation study on the training shift offset $s$. Each curve shows the PSNR on the SIDD Validation dataset as the inference shift offset is varied.}
    \vspace{-3mm}
    \label{fig:fig7}
\end{figure}

\subsection{Ablation Studies}
\label{sec:ablation}

\noindent\textbf{Training shift offset.}
We perform an ablation study to determine the optimal training shift offset \(s\), which is a critical hyperparameter in our network. 
Separate models are trained with \(s \in \{4, 5, 6, 7\}\) and evaluated on the SIDD validation set by varying the inference shift offset, as shown in Fig.~\ref{fig:fig7}.
An offset of \(s=4\) (blue line) is too small to effectively block pixels that are spatially correlated with the target pixel.
This leads to an identity-mapping problem, where the network replicates the noisy input instead of removing noise, resulting in severe PSNR degradation.
Conversely, larger offsets such as \(s=6\) (green line) and \(s=7\) (red line) successfully avoid this issue, but their performance remains suboptimal because an excessively large offset prevents the model from leveraging nearby uncorrelated pixels that provide crucial contextual cues for accurate reconstruction. 
The model trained with \(s=5\) (orange line) achieves the best balance between avoiding identity mapping and utilizing relevant contextual information.
Based on this observation, we select \(s=5\) as the training shift offset for all experiments.

\begin{figure}[t]
    \centering
    \begin{subfigure}[b]{0.32\linewidth}
        \centering
        \includegraphics[width=\linewidth]{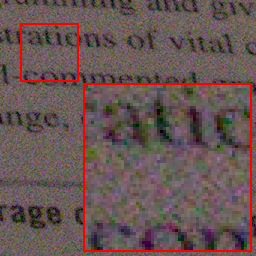}
        \caption{\footnotesize Noisy}
    \end{subfigure}
    \hfill
    \begin{subfigure}[b]{0.32\linewidth}
        \centering
        \includegraphics[width=\linewidth]{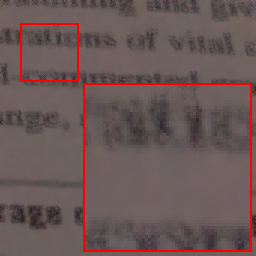}
        \caption{\footnotesize TM-BSN ($s=1$)}
    \end{subfigure}
    \hfill
    \begin{subfigure}[b]{0.32\linewidth}
        \centering
        \includegraphics[width=\linewidth]{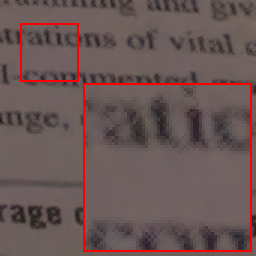}
        \caption{\footnotesize TM-BSN ($s=2$)}
    \end{subfigure}


    \begin{subfigure}[b]{0.32\linewidth}
        \centering
        \includegraphics[width=\linewidth]{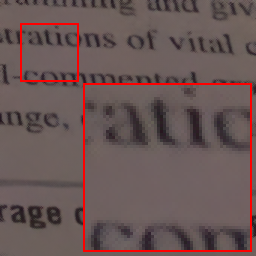}
        \caption{\footnotesize TM-BSN ($s=3$)}
    \end{subfigure}
    \hfill
    \begin{subfigure}[b]{0.32\linewidth}
        \centering
        \includegraphics[width=\linewidth]{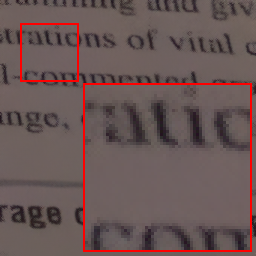}
        \caption{\footnotesize TM-BSN ($s=4$)}
    \end{subfigure}
    \hfill
    \begin{subfigure}[b]{0.32\linewidth}
        \centering
        \includegraphics[width=\linewidth]{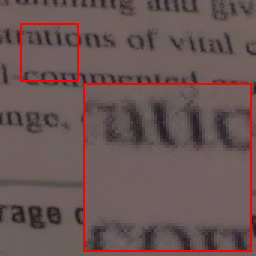}
        \caption{\footnotesize TM-BSN ($s=5$)}
    \end{subfigure}


    \begin{subfigure}[b]{0.32\linewidth}
        \centering
        \includegraphics[width=\linewidth]{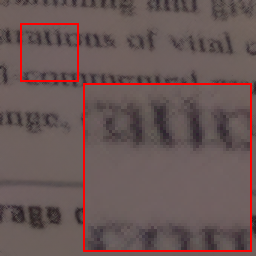}
        \caption{\footnotesize TM-BSN ($s=6$)}
    \end{subfigure}
    \hfill
    \begin{subfigure}[b]{0.32\linewidth}
        \centering
        \includegraphics[width=\linewidth]{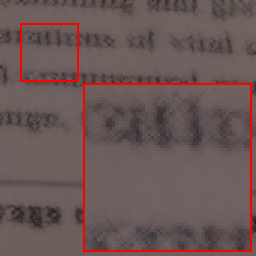}
        \caption{\footnotesize TM-BSN ($s=7$)}
    \end{subfigure}
    \hfill
    \begin{subfigure}[b]{0.32\linewidth}
        \centering
        \includegraphics[width=\linewidth]{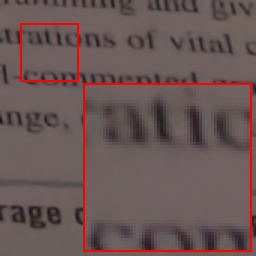}
        \caption{\footnotesize TM-BSN (D)}
    \end{subfigure}

    \caption{Qualitative comparison of denoising results with different shift offsets and the distilled TM-BSN (D).}
    \label{fig:fig8}
\end{figure}

\noindent\textbf{Shift offsets for knowledge distillation.}
We analyze the composition of the shift offset set \(S\), which is used to generate multiple supervision targets for knowledge distillation.
As shown in Table~\ref{tab:table3}, the shift set \(S = \{2, 3, 4, 5, 6\}\) achieves the best quantitative performance, which is consistent with the visual comparisons in Fig.~\ref{fig:fig8}.
Including tiny offsets (e.g., $s=1$) degrades performance because the inference offset deviates substantially from the training offset ($s=5$), causing the teacher to produce unstable supervision signals.
Conversely, using large offsets (e.g., $s\ge7$) limits the model’s ability to utilize the most informative nearby pixels, ultimately leading to degraded texture preservation.
Therefore, we select \(S = \{2, 3, 4, 5, 6\}\) as the optimal configuration to produce diverse yet reliable supervision targets, allowing TM-BSN (D) to achieve further performance gains.

\begin{table}[t]
\centering
\caption{Ablation study on the inference shift offset set $S$ for knowledge distillation. We report PSNR and SSIM on the SIDD validation set. The best results are highlighted in \textbf{bold}.}
\label{tab:table3}
\begin{tabular}{l c c}
\toprule
Shift Set $S$ & PSNR$^\uparrow$(dB) & SSIM$^\uparrow$ \\
\midrule
$\{0, 1, \ldots, 6\}$ & 37.80 & 0.950 \\
$\{0, 1, \ldots, 7\}$ & 37.79 & 0.949 \\
$\{1, 2, \ldots, 6\}$ & 37.81 & 0.951 \\
$\{1, 2, \ldots, 7\}$ & 37.76 & 0.949 \\
$\{2, 3, \ldots, 6\}$ & \textbf{38.08} & \textbf{0.952} \\
$\{2, 3, \ldots, 7\}$ & 37.91 & 0.950 \\
\bottomrule
\end{tabular}
\end{table}

\section{Conclusion}
\label{sec:conclusion}
We have presented TM-BSN, a novel blind-spot architecture that models the spatial correlation of real-world sRGB noise through the proposed triangular-masked convolution.
This design constructs a diamond-shaped blind spot directly at the original image resolution, effectively excluding spatially correlated pixels while fully leveraging uncorrelated contextual information.
We further adopt a knowledge distillation strategy to transfer complementary information from multiple blind-spot predictions, enhancing both accuracy and computational efficiency.
Extensive experiments show that our method achieves state-of-the-art performance on real-world image denoising datasets.

\section*{Acknowledgements}
This work was supported by Samsung Electronics Co., Ltd (IO251211-14315-01), the Institute of Information \& communications Technology Planning \& Evaluation (IITP) grant funded by the Korea government (MSIT) [No. RS-2021-II211343, Artificial Intelligence Graduate School Program (Seoul National University)], and the BK21 FOUR program of the Education and Research Program for Future ICT Pioneers, Seoul National University in 2026.
{
    \small
    \bibliographystyle{ieeenat_fullname}
    \bibliography{main}

@String(CVPR= {IEEE Conf. Comput. Vis. Pattern Recog.})

@String(ICLR = {Int. Conf. Learn. Represent.})

@String(AAAI = {AAAI})

@String(CVPR  = {CVPR})

@String(ICLR  = {ICLR})

@article{n2n,
  title={Noise2Noise: Learning image restoration without clean data},
  author={Lehtinen, Jaakko and Munkberg, Jacob and Hasselgren, Jon and Laine, Samuli and Karras, Tero and Aittala, Miika and Aila, Timo},
  journal={arXiv preprint arXiv:1803.04189},
  year={2018}
}

@inproceedings{n2v,
  title={Noise2void-learning denoising from single noisy images},
  author={Krull, Alexander and Buchholz, Tim-Oliver and Jug, Florian},
  booktitle={Proceedings of the IEEE/CVF conference on computer vision and pattern recognition},
  pages={2129--2137},
  year={2019}
}

@inproceedings{n2s,
  title={Noise2self: Blind denoising by self-supervision},
  author={Batson, Joshua and Royer, Loic},
  booktitle={International conference on machine learning},
  pages={524--533},
  year={2019},
  organization={PMLR}
}

@inproceedings{nbr2nbr,
  title={Neighbor2neighbor: Self-supervised denoising from single noisy images},
  author={Huang, Tao and Li, Songjiang and Jia, Xu and Lu, Huchuan and Liu, Jianzhuang},
  booktitle={Proceedings of the IEEE/CVF conference on computer vision and pattern recognition},
  pages={14781--14790},
  year={2021}
}

@article{hqssl,
  title={High-quality self-supervised deep image denoising},
  author={Laine, Samuli and Karras, Tero and Lehtinen, Jaakko and Aila, Timo},
  journal={Advances in neural information processing systems},
  volume={32},
  year={2019}
}

@inproceedings{dbsn,
  title={Unpaired learning of deep image denoising},
  author={Wu, Xiaohe and Liu, Ming and Cao, Yue and Ren, Dongwei and Zuo, Wangmeng},
  booktitle={European conference on computer vision},
  pages={352--368},
  year={2020},
  organization={Springer}
}

@inproceedings{apbsn,
  title={Ap-bsn: Self-supervised denoising for real-world images via asymmetric pd and blind-spot network},
  author={Lee, Wooseok and Son, Sanghyun and Lee, Kyoung Mu},
  booktitle={Proceedings of the IEEE/CVF Conference on Computer Vision and Pattern Recognition},
  pages={17725--17734},
  year={2022}
}

@article{puca,
  title={Puca: patch-unshuffle and channel attention for enhanced self-supervised image denoising},
  author={Jang, Hyemi and Park, Junsung and Jung, Dahuin and Lew, Jaihyun and Bae, Ho and Yoon, Sungroh},
  journal={Advances in Neural Information Processing Systems},
  volume={36},
  pages={19217--19229},
  year={2023}
}

@inproceedings{lgbpn,
  title={Lg-bpn: Local and global blind-patch network for self-supervised real-world denoising},
  author={Wang, Zichun and Fu, Ying and Liu, Ji and Zhang, Yulun},
  booktitle={Proceedings of the IEEE/CVF conference on computer vision and pattern recognition},
  pages={18156--18165},
  year={2023}
}

@inproceedings{sdap,
  title={Random sub-samples generation for self-supervised real image denoising},
  author={Pan, Yizhong and Liu, Xiao and Liao, Xiangyu and Cao, Yuanzhouhan and Ren, Chao},
  booktitle={Proceedings of the IEEE/CVF international conference on computer vision},
  pages={12150--12159},
  year={2023}
}

@inproceedings{cvfsid,
  title={Cvf-sid: Cyclic multi-variate function for self-supervised image denoising by disentangling noise from image},
  author={Neshatavar, Reyhaneh and Yavartanoo, Mohsen and Son, Sanghyun and Lee, Kyoung Mu},
  booktitle={Proceedings of the ieee/cvf Conference on Computer Vision and Pattern Recognition},
  pages={17583--17591},
  year={2022}
}

@inproceedings{cbsn,
  title={Self-supervised image denoising with downsampled invariance loss and conditional blind-spot network},
  author={Jang, Yeong Il and Lee, Keuntek and Park, Gu Yong and Kim, Seyun and Cho, Nam Ik},
  booktitle={Proceedings of the IEEE/CVF International Conference on Computer Vision},
  pages={12196--12205},
  year={2023}
}

@inproceedings{sasl,
  title={Spatially adaptive self-supervised learning for real-world image denoising},
  author={Li, Junyi and Zhang, Zhilu and Liu, Xiaoyu and Feng, Chaoyu and Wang, Xiaotao and Lei, Lei and Zuo, Wangmeng},
  booktitle={Proceedings of the IEEE/CVF Conference on Computer Vision and Pattern Recognition},
  pages={9914--9924},
  year={2023}
}

@inproceedings{atbsn,
  title={Exploring efficient asymmetric blind-spots for self-supervised denoising in real-world scenarios},
  author={Chen, Shiyan and Zhang, Jiyuan and Yu, Zhaofei and Huang, Tiejun},
  booktitle={Proceedings of the IEEE/CVF Conference on Computer Vision and Pattern Recognition},
  pages={2814--2823},
  year={2024}
}

@inproceedings{aprrd,
  title={APR-RD: Complemental Two Steps for Self-Supervised Real Image Denoising},
  author={Kim, Hyunjun and Cho, Nam Ik},
  booktitle={Proceedings of the AAAI Conference on Artificial Intelligence},
  volume={39},
  number={4},
  pages={4257--4265},
  year={2025}
}

@inproceedings{tbsn,
  title={Rethinking Transformer-Based Blind-Spot Network for Self-Supervised Image Denoising},
  author={Li, Junyi and Zhang, Zhilu and Zuo, Wangmeng},
  booktitle={Proceedings of the AAAI Conference on Artificial Intelligence},
  volume={39},
  number={5},
  pages={4788--4796},
  year={2025}
}

@article{dncnn,
  title={Beyond a gaussian denoiser: Residual learning of deep cnn for image denoising},
  author={Zhang, Kai and Zuo, Wangmeng and Chen, Yunjin and Meng, Deyu and Zhang, Lei},
  journal={IEEE transactions on image processing},
  volume={26},
  number={7},
  pages={3142--3155},
  year={2017},
  publisher={IEEE}
}

@article{ffdnet,
  title={FFDNet: Toward a fast and flexible solution for CNN-based image denoising},
  author={Zhang, Kai and Zuo, Wangmeng and Zhang, Lei},
  journal={IEEE Transactions on Image Processing},
  volume={27},
  number={9},
  pages={4608--4622},
  year={2018},
  publisher={IEEE}
}

@inproceedings{memnet,
  title={Memnet: A persistent memory network for image restoration},
  author={Tai, Ying and Yang, Jian and Liu, Xiaoming and Xu, Chunyan},
  booktitle={Proceedings of the IEEE international conference on computer vision},
  pages={4539--4547},
  year={2017}
}

@inproceedings{sidd,
  title={A high-quality denoising dataset for smartphone cameras},
  author={Abdelhamed, Abdelrahman and Lin, Stephen and Brown, Michael S},
  booktitle={Proceedings of the IEEE conference on computer vision and pattern recognition},
  pages={1692--1700},
  year={2018}
}

@inproceedings{dnd,
  title={Benchmarking denoising algorithms with real photographs},
  author={Plotz, Tobias and Roth, Stefan},
  booktitle={Proceedings of the IEEE conference on computer vision and pattern recognition},
  pages={1586--1595},
  year={2017}
}

@inproceedings{midd,
  title={Real-world mobile image denoising dataset with efficient baselines},
  author={Flepp, Roman and Ignatov, Andrey and Timofte, Radu and Van Gool, Luc},
  booktitle={Proceedings of the IEEE/CVF Conference on Computer Vision and Pattern Recognition},
  pages={22368--22377},
  year={2024}
}

@article{red30,
  title={Image restoration using very deep convolutional encoder-decoder networks with symmetric skip connections},
  author={Mao, Xiaojiao and Shen, Chunhua and Yang, Yu-Bin},
  journal={Advances in neural information processing systems},
  volume={29},
  year={2016}
}

@article{vdir,
  title={Variational deep image restoration},
  author={Soh, Jae Woong and Cho, Nam Ik},
  journal={IEEE Transactions on Image Processing},
  volume={31},
  pages={4363--4376},
  year={2022},
  publisher={IEEE}
}

@inproceedings{cbdnet,
  title={Toward convolutional blind denoising of real photographs},
  author={Guo, Shi and Yan, Zifei and Zhang, Kai and Zuo, Wangmeng and Zhang, Lei},
  booktitle={Proceedings of the IEEE/CVF conference on computer vision and pattern recognition},
  pages={1712--1722},
  year={2019}
}

@article{bm3d,
  title={Image denoising by sparse 3-D transform-domain collaborative filtering},
  author={Dabov, Kostadin and Foi, Alessandro and Katkovnik, Vladimir and Egiazarian, Karen},
  journal={IEEE Transactions on image processing},
  volume={16},
  number={8},
  pages={2080--2095},
  year={2007},
  publisher={IEEE}
}

@inproceedings{wnnm,
  title={Weighted nuclear norm minimization with application to image denoising},
  author={Gu, Shuhang and Zhang, Lei and Zuo, Wangmeng and Feng, Xiangchu},
  booktitle={Proceedings of the IEEE conference on computer vision and pattern recognition},
  pages={2862--2869},
  year={2014}
}

@article{vdn,
  title={Variational denoising network: Toward blind noise modeling and removal},
  author={Yue, Zongsheng and Yong, Hongwei and Zhao, Qian and Meng, Deyu and Zhang, Lei},
  journal={Advances in neural information processing systems},
  volume={32},
  year={2019}
}

@inproceedings{gcbd,
  title={Image blind denoising with generative adversarial network based noise modeling},
  author={Chen, Jingwen and Chen, Jiawei and Chao, Hongyang and Yang, Ming},
  booktitle={Proceedings of the IEEE conference on computer vision and pattern recognition},
  pages={3155--3164},
  year={2018}
}

@inproceedings{c2n,
  title={C2n: Practical generative noise modeling for real-world denoising},
  author={Jang, Geonwoon and Lee, Wooseok and Son, Sanghyun and Lee, Kyoung Mu},
  booktitle={Proceedings of the IEEE/CVF International Conference on Computer Vision},
  pages={2350--2359},
  year={2021}
}

@inproceedings{restormer,
  title={Restormer: Efficient transformer for high-resolution image restoration},
  author={Zamir, Syed Waqas and Arora, Aditya and Khan, Salman and Hayat, Munawar and Khan, Fahad Shahbaz and Yang, Ming-Hsuan},
  booktitle={Proceedings of the IEEE/CVF conference on computer vision and pattern recognition},
  pages={5728--5739},
  year={2022}
}

@inproceedings{malvar,
  title={High-quality linear interpolation for demosaicing of Bayer-patterned color images},
  author={Malvar, Henrique S and He, Li-wei and Cutler, Ross},
  booktitle={2004 IEEE international conference on acoustics, speech, and signal processing},
  volume={3},
  pages={iii--485},
  year={2004},
  organization={IEEE}
}

@inproceedings{unet,
  title={U-net: Convolutional networks for biomedical image segmentation},
  author={Ronneberger, Olaf and Fischer, Philipp and Brox, Thomas},
  booktitle={International Conference on Medical image computing and computer-assisted intervention},
  pages={234--241},
  year={2015},
  organization={Springer}
}

@inproceedings{awgn,
  title={When awgn-based denoiser meets real noises},
  author={Zhou, Yuqian and Jiao, Jianbo and Huang, Haibin and Wang, Yang and Wang, Jue and Shi, Honghui and Huang, Thomas},
  booktitle={Proceedings of the AAAI Conference on Artificial Intelligence},
  volume={34},
  number={07},
  pages={13074--13081},
  year={2020}
}

@article{erf,
  title={Understanding the effective receptive field in deep convolutional neural networks},
  author={Luo, Wenjie and Li, Yujia and Urtasun, Raquel and Zemel, Richard},
  journal={Advances in neural information processing systems},
  volume={29},
  year={2016}
}

@inproceedings{adam,
  author       = {Diederik P. Kingma and
                  Jimmy Ba},
  editor       = {Yoshua Bengio and
                  Yann LeCun},
  title        = {Adam: {A} Method for Stochastic Optimization},
  booktitle    = {3rd International Conference on Learning Representations, {ICLR} 2015,
                  San Diego, CA, USA, May 7-9, 2015, Conference Track Proceedings},
  year         = {2015},
  url          = {http://arxiv.org/abs/1412.6980},
  bibsource    = {dblp computer science bibliography, https://dblp.org}
}

@article{ssim,
  title={Image quality assessment: from error visibility to structural similarity},
  author={Wang, Zhou and Bovik, Alan C and Sheikh, Hamid R and Simoncelli, Eero P},
  journal={IEEE transactions on image processing},
  volume={13},
  number={4},
  pages={600--612},
  year={2004},
  publisher={IEEE}
}

@ARTICLE{pcsd,
  author={Xiaolin Wu and Ning Zhang},
  journal={IEEE Transactions on Image Processing}, 
  title={Primary-consistent soft-decision color demosaicking for digital cameras (patent pending)}, 
  year={2004},
  volume={13},
  number={9},
  pages={1263-1274},
  doi={10.1109/TIP.2004.832920}}

@ARTICLE{gunturk,
  author={Gunturk, B.K. and Altunbasak, Y. and Mersereau, R.M.},
  journal={IEEE Transactions on Image Processing}, 
  title={Color plane interpolation using alternating projections}, 
  year={2002},
  volume={11},
  number={9},
  pages={997-1013},
  doi={10.1109/TIP.2002.801121}}

@ARTICLE{zhang,
  author={Lei Zhang and Xiaolin Wu},
  journal={IEEE Transactions on Image Processing}, 
  title={Color demosaicking via directional linear minimum mean square-error estimation}, 
  year={2005},
  volume={14},
  number={12},
  pages={2167-2178},
  doi={10.1109/TIP.2005.857260}}

@article{tv,
  title={Nonlinear total variation based noise removal algorithms},
  author={Rudin, Leonid I and Osher, Stanley and Fatemi, Emad},
  journal={Physica D: nonlinear phenomena},
  volume={60},
  number={1-4},
  pages={259--268},
  year={1992},
  publisher={Elsevier}
}

@inproceedings{nlm,
  title={A non-local algorithm for image denoising},
  author={Buades, Antoni and Coll, Bartomeu and Morel, J-M},
  booktitle={2005 IEEE computer society conference on computer vision and pattern recognition (CVPR'05)},
  volume={2},
  pages={60--65},
  year={2005},
  organization={Ieee}
}

@article{polyu,
  title={Real-world noisy image denoising: A new benchmark},
  author={Xu, Jun and Li, Hui and Liang, Zhetong and Zhang, David and Zhang, Lei},
  journal={arXiv preprint arXiv:1804.02603},
  year={2018}
}

@inproceedings{cycleisp,
  title={Cycleisp: Real image restoration via improved data synthesis},
  author={Zamir, Syed Waqas and Arora, Aditya and Khan, Salman and Hayat, Munawar and Khan, Fahad Shahbaz and Yang, Ming-Hsuan and Shao, Ling},
  booktitle={Proceedings of the IEEE/CVF conference on computer vision and pattern recognition},
  pages={2696--2705},
  year={2020}
}

@inproceedings{n2nf,
  title={Noise2noiseflow: Realistic camera noise modeling without clean images},
  author={Maleky, Ali and Kousha, Shayan and Brown, Michael S and Brubaker, Marcus A},
  booktitle={Proceedings of the IEEE/CVF Conference on Computer Vision and Pattern recognition},
  pages={17632--17641},
  year={2022}
}

@inproceedings{neca,
  title={sRGB real noise synthesizing with neighboring correlation-aware noise model},
  author={Fu, Zixuan and Guo, Lanqing and Wen, Bihan},
  booktitle={Proceedings of the IEEE/CVF Conference on Computer Vision and Pattern Recognition},
  pages={1683--1691},
  year={2023}
}
}

\end{document}